\documentclass[letterpaper, 10 pt, conference]{ieeeconf}  

\IEEEoverridecommandlockouts                              

\usepackage{amsmath,amsfonts}
\usepackage{algorithmic}
\usepackage{algorithm}
\usepackage{array}
\usepackage[caption=false,font=normalsize,labelfont=sf,textfont=sf]{subfig}
\usepackage{textcomp}
\usepackage{stfloats}
\usepackage{url}
\usepackage{verbatim}
\usepackage{graphicx}
\usepackage{cite}
\usepackage{bm}
\usepackage{color}
\usepackage{booktabs}
\usepackage{multirow}
\usepackage{makecell}
\usepackage{graphicx}
\usepackage{soul}
\usepackage{xcolor}

\usepackage{glossaries}
\makeglossaries
\newacronym{strmf}{SRMF}{Smoothness Regularized Matrix Factorization}

\title{\LARGE \bf
High-Dimensional Fault Tolerance Testing of Highly Automated Vehicles Based on Low-Rank Models
}

\author{Yuewen Mei,
        Tong Nie,~\IEEEmembership{Student Member,~IEEE,}
        Jian Sun,
        Ye Tian,~\IEEEmembership{Member,~IEEE}
\thanks{*Research supported by the National Key Research and Development Program of China under Grant [2021YFB2501202], and the National Natural Science Foundation of China under Grant [52172391].}
\thanks{The authors are with the Department of Traffic Engineering and Key Laboratory of Road and Traffic Engineering, Ministry of Education, Tongji University. Shanghai, China. 201804. (E-mail: meiyuewen@tongji.edu.cn; nietong@tongji.edu.cn; sunjian@tongji.edu.cn; tianye@tongji.edu.cn)}%
\thanks{Corresponding author: Ye Tian (tianye@tongji.edu.cn)}%
}

\begin{document}

\maketitle
\thispagestyle{empty}
\pagestyle{empty}
\begin{abstract}
Ensuring fault tolerance of Highly Automated Vehicles (HAVs) is crucial for their safety due to the presence of potentially severe faults. Hence, Fault Injection (FI) testing is conducted by practitioners to evaluate the safety level of HAVs. To fully cover test cases, various driving scenarios and fault settings should be considered. However, due to numerous combinations of test scenarios and fault settings, the testing space can be complex and high-dimensional. In addition, evaluating performance in all newly added scenarios is resource-consuming. The rarity of critical faults that can cause security problems further strengthens the challenge.
To address these challenges, we propose to accelerate FI testing under the low-rank Smoothness Regularized Matrix Factorization (SRMF) framework. We first organize the sparse evaluated data into a structured matrix based on its safety values. Then the untested values are estimated by the correlation captured by the matrix structure. 
To address high dimensionality, a low-rank constraint is imposed on the testing space. 
To exploit the relationships between existing scenarios and new scenarios and capture the local regularity of critical faults, 
three types of smoothness regularization are further designed as a complement.
We conduct experiments on car following and cut in scenarios. The results indicate that SRMF has the lowest prediction error in various scenarios and is capable of predicting rare critical faults compared to other machine learning models. In addition, SRMF can achieve 1171 acceleration rate, 99.3\% precision and 91.1\% F1 score in identifying critical faults. To the best of our knowledge, this is the first work to introduce low-rank models to FI testing of HAVs.
\end{abstract}


\section{INTRODUCTION}
Highly Automated Vehicles (HAVs) can greatly improve the safety, mobility, and efficiency of transportation \cite{feng2023dense}. However, ensuring the safety performance and resilience of HAVs has been a long-standing concern due to fatal accidents caused by HAVs. HAVs can be vulnerable and dangerous when exposed to faults caused by malicious attackers \cite{hoque2022autonomous} or system malfunctions \cite{jha2019kayotee}. Malicious attackers are interested in creating a potential accident by tampering with the data in HAVs without authorization. For example, corrupting the hardware or software to force the HAVs to deviate from the lane \cite{jha2020ml}. Meanwhile, system malfunctions are frequently posed in complex HAVs. These minor faults can lead to severe consequences, such as collisions with obstacles. Hence, Fault Injection (FI) testing becomes a necessary verification and validation methods for HAVs \cite{ma_verification_2022}.


\begin{figure}[!t]
\centering
  \includegraphics[width=0.45\textwidth]{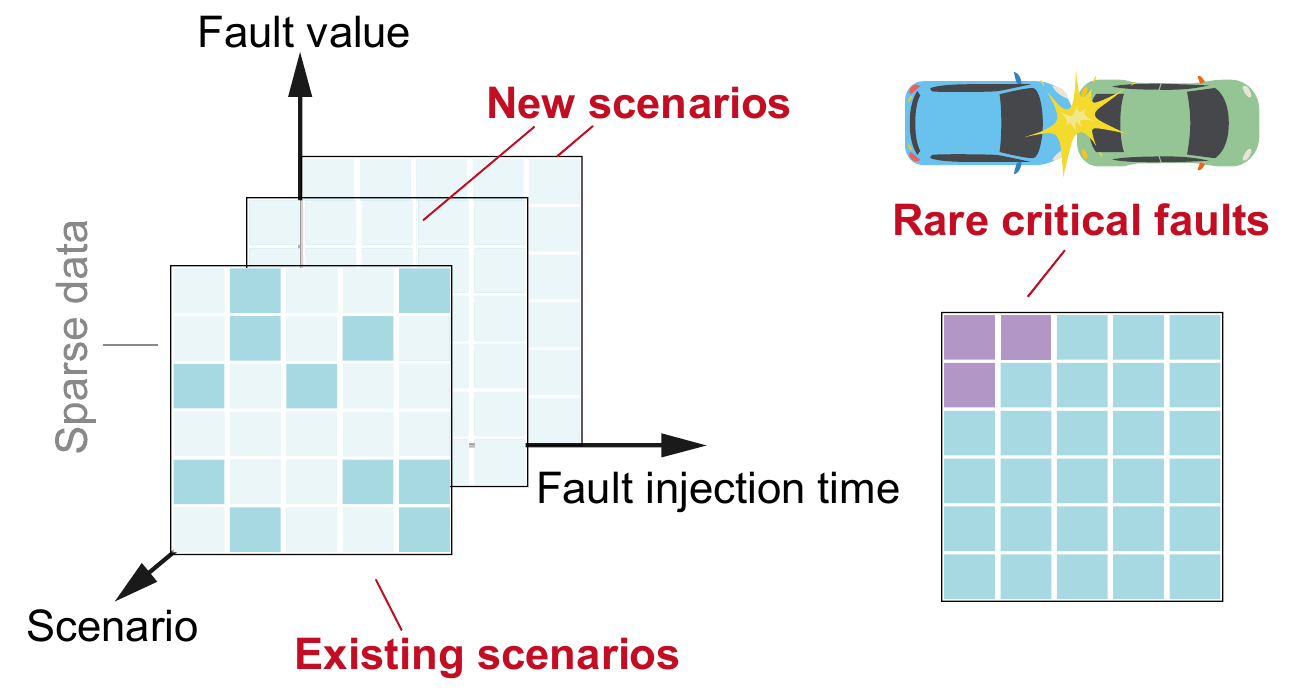}
  \caption{Illustration of fault injection testing problem. The high dimensional testing space consists of scenario parameters and fault parameters. Only a few testing results can be obtained, which compose the sparse data. However, predicting new scenarios and rare critical faults may be difficult.}
  \label{problem}
\end{figure}

In reality, HAVs face a variety of complex driving scenarios. Due to the complexity and randomness, a fault can occur in an arbitrary location and time.
Accordingly, the scenario space and the fault space are intermixed to form a large-scale space to be tested, leading to high-dimensional problems. To avoid the serious consequences of FI testing in the physical world, testing is usually carried out based on simulation. Due to constraints on testing resources, only a limited number of simulation testing can be conducted and these simulations constitute the sparse simulation data. 
Consequently, many accelerated testing methods are proposed to predict the performances in those untested cases and find critical scenarios or critical faults that can have a high potential to cause security problems \cite{Sun2022scenario-based}. Among these accelerated testing methods, surrogate models, such as Gradient Boosted Trees \cite{chen_adaptive_2023}, are commonly applied to fit the evaluated data points and directly predict the testing results given new input testing tasks, bypassing the need to conduct time-consuming simulations per case.

However, although existing surrogate models have demonstrated effectiveness in some testing settings \cite{sun2021adaptive}, there are still problems that need to be further addressed. 
First, the combination of scenario space and fault space generates a mixed testing space with high-dimensional correlations between elements. Commonly used surrogate models such as XGB treat each element as an independent sample and ignore the correlation.
Second, real-world testing tasks often require the evaluation of newly added scenarios. To reduce resource consumption, surrogate models need to directly predict the performance in new scenarios based on the knowledge learned from historical scenarios. However, many popular surrogate methods are prone to overfit the training scenario sets, leading to poor performance in new scenarios without retraining. 
Third, critical faults that can induce safety violations of HAVs are usually rare in the entire space. This rarity is usually manifested as some local patterns in the testing space. Surrogate models that do not consider the local regularities may fail to find all critical faults thoroughly. 

To address these limitations, in this work, we propose to accelerate the FI testing with a simple-yet-effective method, called \textit{low-rank} model. Low-rank models are widely adopted to analyze high-dimensional arrays with sparse observations. 
In this framework, sparse observation data are ordered as incomplete matrices or tensors. The complex dependencies and correlations between different dimensions are then captured by the structure of the matrix and tensor \cite{liu2012tensor}. Due to the low-rank assumption, it is particularly efficient for reconstructing high-dimensional correlated data from sparse observations.
For example, it can help optimize the hyperparameter in machine learning algorithms from high-dimensional search spaces \cite{deng2022new}, and estimate the traffic speed from missing data based on spatiotemporal correlation \cite{nie2023correlating}. 

To adapt the standard low-rank models for FI testing, we first characterize the simulation data for FI testing as high-dimensional, sparse data with inherent correlations. To address the aforementioned three challenges in FI testing, we formulate the FI testing problem as a low-rank matrix completion problem. The sparse evaluated results are organized as a fault matrix, where the unknown results of new scenarios and new faults are predicted by the correlation in this matrix. To exploit the high-dimensional correlation of scenarios and fault parameters, we capture the pattern of the entire fault matrix from the low-rank property. To predict newly added scenarios without any historical value, we apply smoothness regularization across different scenarios. The consistency between scenarios can be trained from existing scenarios and then used to predict new scenarios. To handle the rarity problem in some scenarios, local regularization applied in adjacent fault parameters is designed. We term the proposed model as \gls{strmf}. 
Finally, we conduct experiments based on car-following and cut-in scenarios to verify the accuracy and efficiency of \gls{strmf}. 
To the best of our knowledge, this is the first work to introduce low-rank models to FI testing of HAVs.
The main contributions of this work are summarized as follows:
\begin{itemize}
\item We are the first to propose to accelerate the FI testing under limited testing resources with a low-rank matrix factorization model that considers high-dimensional correlations of testing space. 
\item To predict the new untested scenarios and better locate the critical faults, we design three types of local smoothness regularization and integrate them into the matrix factorization framework.
\end{itemize}


The rest of this paper is organized as follows. Section II provides recent related works on FI testing, surrogate models, and low-rank models. Section III formulates the FI problem. Section IV describes the data organization of FI and explains the methodology of \gls{strmf}. Section V conducts an experiment to verify the effectiveness of \gls{strmf}. Section VI concludes the work and discusses future work.

\section{Related works}
Safety testing is crucial for HAVs before large-scale deployment. There are three verification and validation methods for HAV safety testing: scenario-based testing, formal verification, and FI testing \cite{ma_verification_2022}. However, testing scenarios and faults consist of a high-dimensional space and randomly selecting scenarios or faults for testing is inefficient \cite{Moradi2020}. Accordingly, to accelerate the HAVs safety testing process, surrogate models have been widely used. 

Surrogate models can approximate the testing results, thus saving tremendous testing resources, such as searching for critical faults \cite{chen_adaptive_2023, mei2024}, finding faulty behavior from the parameter space \cite{Beglerovic2017}, identifying high-risk scenarios \cite{sun2021adaptive}, and discovering safety performance boundaries \cite{MULLINS2018197, wang2022safety}. Based on the summary of the machine learning methods commonly used as surrogate models, \cite{angione2022using} compares the strengths and weaknesses of various surrogate models. However, the comparison clearly shows that Decision Trees, Random Forrest, Gradient Boosted Trees (such as Extreme Gradient Boosting), and Neural Networks are prone to overfitting. This leads to the surrogate models performing poorly in new scenarios that have not been tested before. 

Low-rank models are widely adopted to reconstruct high-dimensional data with low-dimensional structures. This kind of model assumes the algebraic structures of the observations and can reconstruct the unobserved data in a self-supervised way, without the need to collect a large amount of labeled training data.  
For example, visual data can be organized into a three-dimensional tensor, and missing pixels can be recovered with a low-rank solution. \cite{liu2012tensor}. In the field of spatiotemporal traffic data, low-rank models garnered great interest in missing data imputation problems. Representative methods include tensor factorization \cite{tan2013tensor}, nuclear norm minimization \cite{nie2022truncated}, and matrix factorization \cite{asif2016matrix}. In addition, they can be integrated with advanced machine learning frameworks to tackle more challenging tasks, such as deep learning architectures \cite{yang2021real,nie2023imputeformer,nie2024spatiotemporal} and surrogate modeling \cite{luise2019leveraging}.

\section{Problem Formulation}
This paper considers three types of scenarios: functional scenario, logical scenario, concrete scenario in the scenario-based testing \cite{PEGASUS2017Scenario}. The functional scenario describes the participants and their behaviors within the scenario. The logical scenario parameterizes the scenario, by giving the definition, distribution, and range of the parameters. Then, in the concrete scenario, these parameters are precisely defined. We denote $k$ types of functional scenarios as $P_1, P_2, \ldots, P_{k}$ and each functional scenario $P_i$ has a parameter space containing $n_i$ parameters, which correspond to $n_i$ types of concrete scenarios denoted as $p_{i,n_i}$. Then we have:

\begin{equation}
\begin{aligned}
    &|P_i| = n_i, \forall i = 1,2,\ldots,k \\
    &P_i = \{ p_{i,1}, p_{i,2}, \ldots, p_{i,n_i} \}
\end{aligned}
\end{equation}

Therefore, there are $K$ concrete scenarios to be tested under fault injection in total.
\begin{equation}
    K = \sum_i^{k} n_i
\end{equation}

The fault space in a concrete scenario can consist of the fault value, the injection time, the duration and the location of the fault \cite{xing_adaptive_2023, Jha2019ML-Based,chen_adaptive_2023}. In this work, two fault parameters are considered: fault value and fault injection time, because their values are more diverse and form a higher-dimensional fault space denoted by a matrix $\boldsymbol{F}$. The number of types of fault value and fault injection time is denoted as $I,J$ respectively.
\begin{equation}
    \boldsymbol{F} = \left[\begin{array}{cccc}
    f_{00} & f_{01} & \ldots  & f_{0J} \\
    f_{10} & f_{11} & \ldots  & f_{1J} \\
    \ldots & \ldots & \ldots & \ldots \\
    f_{I0} & f_{I1} & \ldots  & f_{IJ} \\
    \end{array}\right]
\end{equation}

Critical faults are defined as faults that result in severe consequences, such as a collision with other vehicles. The safety impact of the fault $f_{ij}$ is marked as the safety indicator $x_{ij}$. $x_{ij}$ equals to the negative value of Time To Collision (TTC) when no collision occurs and the severity of the collision when a collision occurs \cite{chen_adaptive_2023}. Therefore, all critical faults $F_{\text{critical}}$ can be marked as the set of all faults with a safety indicator larger than 0.
\begin{equation}
    F_{\text{critical}} = \{f_{ij}|x_{ij} > 0\}
\end{equation}

To thoroughly test the fault tolerance of HAVs, it is important to find $F_{\text{critical}}$ in the $K$ scenarios. 

\section{Methodology}
\subsection{Data organization}
The $K$ scenarios can be divided into partially tested scenarios (existing scenarios) and completely new scenarios. In the concrete scenarios of functional scenario $P_i$, suppose that the proportion of faults that have been tested to the entire fault space is $r_i$. Meanwhile, only part of the concrete scenarios in $P_i$ are tested, and others are not tested at all. The existing concrete scenarios are distributed at equal intervals. For example, with an interval of 3, the proportion of tested faults in different concrete scenarios is:
\begin{equation}
    (r_{i,1}, r_{i,2}, r_{i,3}, r_{i,4},r_{i,5}, \ldots) = (r_i,0,0,r_i,0,\ldots)
\end{equation}

\begin{figure*}[!tb]
\centering
  \includegraphics[width=1\textwidth]{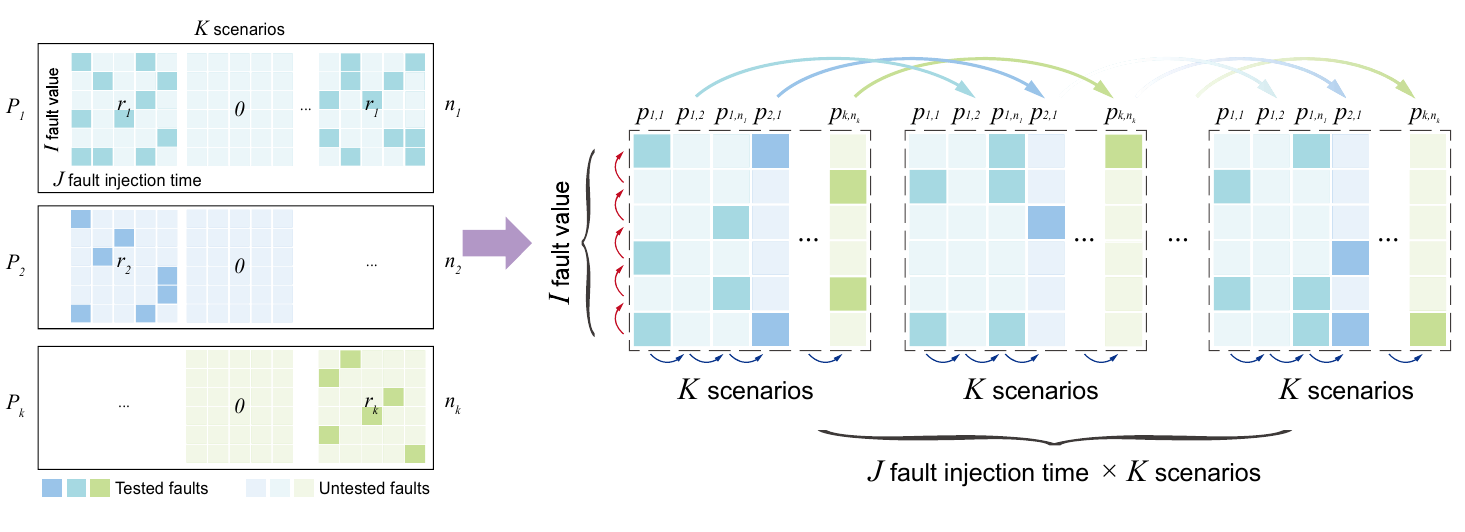}
  \caption{Data organization and smoothness regularization. (1) There are $K$ concrete scenarios to be tested in total, with $P_1, P_2, ...,P_k$ functional scenarios marked in different colors. In each concrete scenario, there are $I$ types of fault value and $J$ types of fault injection time. Only a subset of concrete scenarios and only a subset of faults are tested, with the darker colors indicating the tested ones. (2) Then, the data is transformed into a matrix $\boldsymbol{X}\in\mathbb{R}^{(JK)\times I}$. Three kinds of smoothness regularization are applied. The red arrows represent the smoothness regularization in rows, connecting different fault value. The dark blue arrows and the arrows drawn at the top are used in columns to capture the relationship between scenarios and fault injection time.}
  \label{Methodology}
\end{figure*}

To capture the relationship between existing scenarios and new scenarios from the two fault dimensions, the $IJ$ faults injected into the $K$ scenarios are folded as illustrated in Fig. \ref{Methodology}. 
The scenarios are arranged by one fault parameter: fault injection time. For example, in the first part of the folded matrix, the fault injection time is the same while there are $K$ types of scenarios with $I$ types of fault value. 
Therefore, the safety indicators of different scenarios and faults can be formed as a matrix $\boldsymbol{X}\in\mathbb{R}^{I\times (JK)}$. The rows correspond to the $I$ types of fault value, and the columns correspond to $K$ scenarios with $J$ types of injection time.

The index set of the scenarios and faults that have been tested in $\boldsymbol{X}$ is denoted as $\Omega$. Therefore, the observed values in the matrix can be represent by a sampling operator: $\mathcal{P}_{\Omega}(\boldsymbol{X})$:  
\begin{equation}
    [\mathcal{P}_{\Omega}(\boldsymbol{X})]_{i,m}=\left\{\begin{array}{ll}
    y_{i,m},     & \text{if $(i,m)\in\Omega$,}  \\
    0,     & \text{otherwise}, \\
    \end{array}\right.
\end{equation}
where $i=1,\ldots,I$ and $m=1,\ldots,JK$.

\subsection{Smoothness regularized matrix factorization}
In standard Low-Rank Matrix Factorization (MF) method, $\boldsymbol{X}$ is calculated by factorizing it into two low-rank factor matrices: $\boldsymbol{W}\in\mathbb{R}^{R\times I}$ for rows and $\boldsymbol{H}\in\mathbb{R}^{R\times M}$ for columns, where $M = JK, R\leq \min(I, M)$. Then, we can estimate the complete matrix $\boldsymbol{X}$ from the low-rank matrix $\boldsymbol{W}^\top \boldsymbol{H}$ with a reconstruction loss on the observed points:
\begin{equation}
    \min _{\boldsymbol{W}, \boldsymbol{H}} \frac{1}{2}\|\mathcal{P}_{\Omega}(\boldsymbol{X} - \boldsymbol{W}^\top \boldsymbol{H})\|_{F}^2 + \frac{\rho}{2}(\|\boldsymbol{W}\|_{F}^2 + \|\boldsymbol{H}\|_{F}^2)
\label{mf}
\end{equation}
where $\rho$ is the weight parameters for regularization terms.

The rationale for low-rank models is to reconstruct the dominating features of the space.
Both the scenario parameters and the fault parameters change successively. The safety indicators obtained by simulation testing with adjacent parameters are similar to each other. Then, the correlation in the high-dimensional testing space is significant. 
Therefore, low-rank assumptions can encourage the estimation of dominating patterns of fauthuse, thereby predicting the safety values of untested faults.

The MF model can capture the global pattern of the whole fault matrix. However, the local correlation within fault values (between adjacent rows) and fault injection time, scenarios (between adjacent columns) are neglected. To describe the local correlation, we add smoothness regularization and auto-regression in MF.

The fault injection time ranges from the beginning to the end of the scenario at a fixed time step, and the scenario parameters is similar in adjacent scenarios. The safety indicators are smoothly varied from one parameter to another.
Therefore, in the columns representing scenarios and fault injection times, two smoothness regularization are designed to preserve the continuity between adjacent scenarios and adjacent fault injection time separately. We first formulate two difference operators as follows:
\begin{equation}
    \boldsymbol{\Psi}_{r1} = \left[ \boldsymbol{0}_{(M-1)\times 1} \boldsymbol{I}_{M-1} \right] - \left[ \boldsymbol{I}_{M-1} \boldsymbol{0}_{(M-1)\times 1} \right] 
\end{equation}
\begin{equation}
    \boldsymbol{\Psi}_{r2} = \left[ \boldsymbol{0}_{(M-K)\times K} \boldsymbol{I}_{M-K} \right] - \left[ \boldsymbol{I}_{M-K} \boldsymbol{0}_{(M-K)\times K} \right]
\end{equation}
where $\Psi_{r1}$ represents the index of two adjacent scenarios with the same fault injection time.  $\Psi_{r2}$ signifies the index of two adjacent fault injection times of a single scenario. 

Accordingly, the difference of safety values in $\boldsymbol{H}$ can be formed as:
\begin{equation}\label{eq:smoothness1}
\boldsymbol{H}\boldsymbol{\Psi}_{r1}^\top = \left[ \begin{array}{ccc}
\mid &  &\mid \\
\boldsymbol{h}_2-\boldsymbol{h}_1 & \ldots & \boldsymbol{h}_M-\boldsymbol{h}_{M-1} \\
\mid &  &\mid \\
 \end{array}\right]
\end{equation}
\begin{equation}\label{eq:smoothness2}
\boldsymbol{H}\boldsymbol{\Psi}_{r2}^\top = \left[ \begin{array}{ccc}
\mid &  &\mid \\
\boldsymbol{h}_K-\boldsymbol{h}_1 & \ldots & \boldsymbol{h}_M-\boldsymbol{h}_{M-K} \\
\mid &  &\mid \\
 \end{array}\right]
\end{equation}

In the rows representing fault values, the auto-regression is applied to consider the index-lagged correlation between fault values in $\boldsymbol{W}$:
\begin{equation}
    \boldsymbol{w}_i = \sum_{u=1}^{l}\boldsymbol{T}_u \boldsymbol{w}_{i-u} + \boldsymbol{\epsilon}_i , i = l+1, l+2,\ldots,I
\end{equation}
where $\boldsymbol{w}_i$ is the $i$-th vector in $\boldsymbol{W}$, $\boldsymbol{T}_u$ is the coefficient matrix of auto-regression, $l$ is the order of auto-regression, $\boldsymbol{\epsilon}_i$ is the residual vector.
\begin{equation}
    \boldsymbol{\Psi}_u = \left[ \boldsymbol{0}_{(I-l)\times(l-u)}  \boldsymbol{I}_{I-l}  \boldsymbol{0}_{(I-l)\times u} \right] 
\end{equation}

Then, the auto-regression between variates can be formulated as: 
\begin{equation}\label{eq:autoreg}
    \boldsymbol{W}_i\boldsymbol{\Psi}^\top_0 \approx \sum_{u=1}^{l}\boldsymbol{T}_u\boldsymbol{W}\boldsymbol{\Psi}^\top_u 
\end{equation}

Therefore, the row- and column-wise smoothness regularization is added to Eq. \ref{mf}:
\begin{equation}\label{eq:overall}
\begin{aligned}
    \min _{\boldsymbol{W}, \boldsymbol{H}, \boldsymbol{T}_u} & \frac{1}{2}\|\mathcal{P}_{\Omega}(\boldsymbol{X} - \boldsymbol{W}^\top \boldsymbol{H})\|_{F}^2 + \frac{\rho}{2}(\|\boldsymbol{W}\|_{F}^2 + \|\boldsymbol{H}\|_{F}^2) \\ 
    &+\frac{\lambda_1}{2}\|\boldsymbol{H}\Psi_{r1}^\top\|_{F}^2 +\frac{\lambda_2}{2}\|\boldsymbol{H}\Psi_{r2}^\top\|_{F}^2 \\ 
    &+\frac{\lambda_3}{2}\|\boldsymbol{W}\Psi^\top_0-\sum_{u=1}^{l} \boldsymbol{T}_u\boldsymbol{W}\Psi^\top_u\|_{F}^2,
\end{aligned}
\end{equation}
where $\lambda_1$, $\lambda_2$, $\lambda_3$ are the coefficients of the row and column smoothness regularization respectively.

Eqs. \eqref{eq:smoothness1}, \eqref{eq:smoothness2} and \eqref{eq:autoreg} regularizes the local continuity of the fault space, serving as a complement to the global low-rank property of MF. This feature enables our model to address the newly added scenario problem and the rare critical fault problem.

The convex optimization problem in Eq. \eqref{eq:overall} can be solved by the alternating minimization algorithm \cite{yu2016temporal,chen2022nonstationary}. Then, the complete fault matrix $\boldsymbol{X}$ can be obtained to estimate the safety indicator of the untested scenarios and faults, thus saving the time cost of conducting the simulation.


\section{Case study}

\subsection{Experimental settings}
\begin{figure}[!htpb]
    \centering
    \includegraphics[width = 0.45\textwidth]{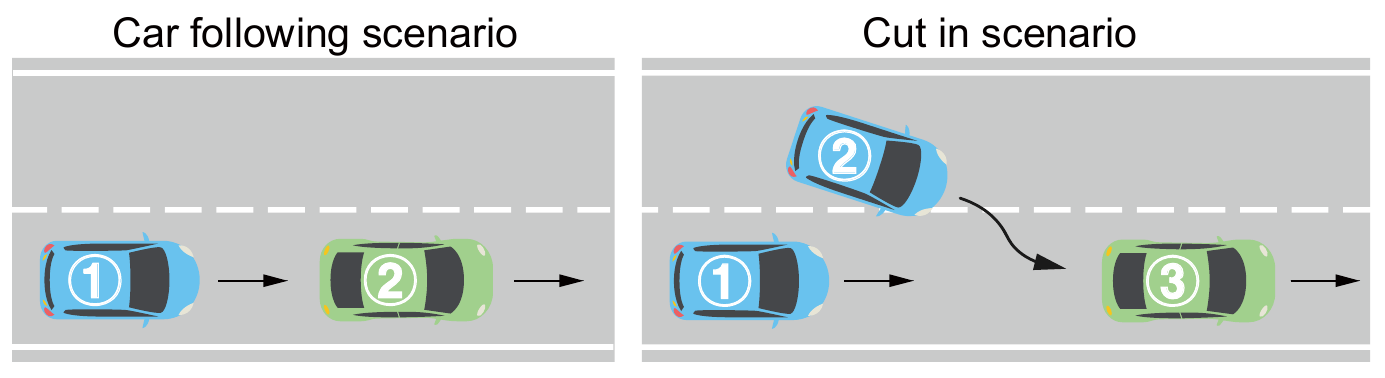}
    \caption{Car following and cut in scenario.}
    \label{fig:scenario}
\end{figure}

To verify the effectiveness of \gls{strmf}, two functional scenarios are tested: cut in (denoted as $P_1$) and car following (denoted as $P_2$) in our experiment, as shown in Fig. \ref{fig:scenario}. In the car following scenario, the blue vehicle 1 behind controlled by HAV follows the leading vehicle 2. While in the cut in scenario, vehicle 2 controlled by HAV in the left lane preparing to change lanes in front of vehicle 1. 

The HAVs are controlled with the Intelligent Driver Model (IDM) and the IDM parameter is the same as that calibrated and defined in \cite{sun2021adaptive}.


\begin{figure*}[!thpb]
\centering
  \includegraphics[width=0.98\textwidth]{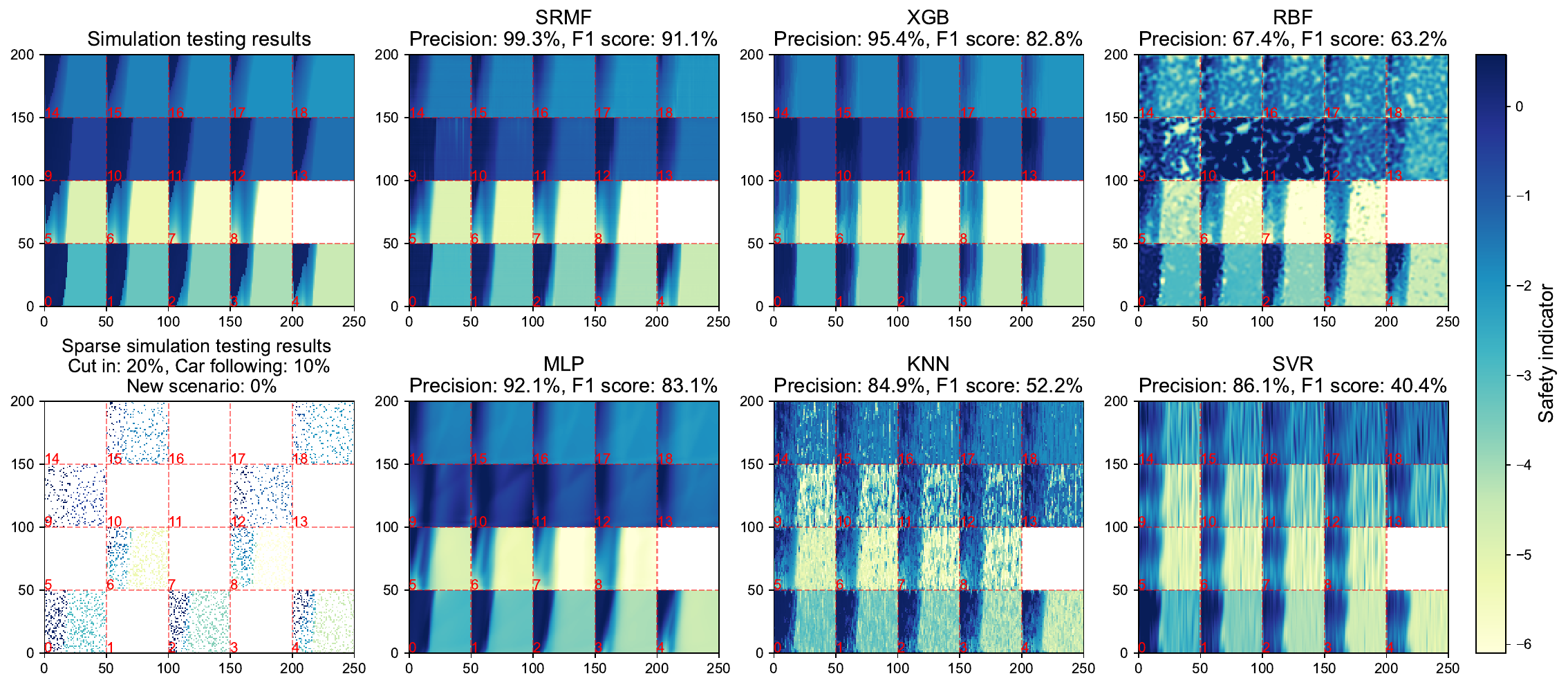}
  \caption{Simulation and prediction results of cut in and car following scenario. The number colored in red are the scenario id and each grid divided by red dotted lines represents a scenario with $50*50$ kinds of faults. The X-axis represents fault injection time and Y-axis represents fault value. Safety indicator of each fault is shown by the color. The darker the color, the more serious the fault. The precision of each model is also marked.}
  \label{simulation_and_prediction_results}
\end{figure*}

The fault space and the concrete scenario space are shown in TABLE \ref{space}, that is, $I = 50, J = 50, n_1 = 9, n_2 = 10, K = 19$. The initial distance is the longitudinal distance between vehicle 1 and vehicle 2. Other scenario parameters are set to fixed values in this experiment. The interval of existing concrete scenarios is 2 in cut in scenario and 3 in car following scenario. Meanwhile, the tested faults proportion is set as: $r_1 = 0.2, r_2 = 0.1$. The simulation testing results of the entire testing space, the sparse simulation testing results with a few tested faults are depicted in the left side of Fig. \ref{simulation_and_prediction_results}. Scenario from \# $0$ to \# $8$ are cut in scenario and scenario from \# $9$ to \# $18$ are car following scenario. Scenario \# $1, 3, 5, 7, 10, 11, 13, 14, 16, 17$ are newly added scenarios without any simulations. Scenarios \# $0, 2, 4, 6, 8$ are the tested cut in scenarios with $20\%$ faults tested and scenarios \# $9, 12, 15, 18$ are the tested car following scenarios with $10\%$ faults tested.

The hyper parameters in \gls{strmf} are specified as follows. ${\lambda_1}$ is set to 1, ${\lambda_2}$ is set to 1, and ${\lambda_3}$ is set to 10. ${\rho}$ is set to 0.01 and the rank of the matrix is 10. We set the maximum number of iterations as 150.

We take other surrogate models used in HAVs safety testing\cite{sun2021adaptive,chen_adaptive_2023,wang2022safety}, including Multi-layer Perception (MLP), Extreme Gradient Boosting (XGB), K-Nearest Neighbor Regression (KNN), Radial Basis Functions (RBF), Support Vector Regression (SVR) as the baseline models. The structure and hyper parameters of MLP is the same as that in \cite{wang2022safety}. The hyper parameters of XGB, KNN, SVR, RBF are adjusted according to \cite{sun2021adaptive}. The prediction error, including the Mean Absolute Error (MAE), the Weighted Mean Absolute Percentage Error (WMAPE), of safety indicator $\boldsymbol{X}$ and precision, F1 score of critical faults $F_{\text{critical}}$ are used as evaluation metrics.


\begin{table}[h]
\caption{Fault space and scenario space tested in this experiment}
\fontsize{8pt}{10pt}\selectfont
    \centering
    \resizebox{0.98\columnwidth}{!}{
    
    \begin{tabular}{c|cc|cc}
    \hline
            &   \multicolumn{2}{c}{Fault space}&	\multicolumn{2}{c}{Scenario space}\\
            &\makecell{Injection\\ time \\ (time step)}&\makecell{Fault\\value\\(m/s\textsuperscript{2})}& \makecell{Cut in\\initial distance \\ (m)}& \makecell{Car following\\initial distance \\ (m)} \\
    \hline
        Minimum &0	&0	&5	&16\\
        Maximum &49	&4.9	&13	&25\\
        Intervals &1	&0.1	&1	&1 \\
        Number &50 & 50 & 9 & 10\\
    \hline
    \end{tabular}
    }
    \label{space}
\end{table}

\subsection{Acceleration Rate of \gls{strmf}}
The time cost to simulate the untested car following and cut in testing space (44,462 types of scenarios and faults combinations) is 1875 s. In comparison, it only takes 1.6 s for \gls{strmf} to calculate these results. The acceleration rate of \gls{strmf} can reach 1171 times. 

\subsection{Effectiveness of \gls{strmf} in new scenarios}
The prediction results of the six models are shown in Fig. \ref{simulation_and_prediction_results}. Performance evaluations for these six models are shown in TABLE \ref{metrics_of_all_scenario}. During prediction, only sparse simulation testing results are taken as input and the safety indicator values are transferred to positive values for matrix factorization. 
\begin{table}[h]
\caption{Prediction error and Precision}
\fontsize{7.2pt}{11pt}\selectfont
\vspace{-10pt}
\label{metrics_of_all_scenario}
\begin{center}
\resizebox{0.98\columnwidth}{!}{
\begin{tabular}{cc|cccccc}
\hline
\multicolumn{2}{c}{Metrics} & \gls{strmf} &MLP & XGB & KNN & RBF  &SVR\\
\hline
\multirow{4}{*}{\makecell{All\\scenarios}}&MAE & \textbf{0.074} &  0.226&	0.215&	0.830&	0.734	&0.993\\
&WMAPE	&\textbf{0.018}	&0.056	&0.053&	0.206&	0.182&	0.246\\
&Precision	&\textbf{0.993}	&0.921&	0.954&	0.849	&0.674&	0.861\\
&F1 score	&\textbf{0.911}	&0.831&	0.828&	0.522&	0.632&	0.404\\
\hline
\multirow{4}{*}{\makecell{Existing\\scenarios}}&MAE & \textbf{0.055}	&0.191	&0.122&	0.859&	0.502	&1.009\\
&WMAPE	&\textbf{0.014}	&0.049	&0.032&	0.222&	0.130	&0.261\\
&Precision	&\textbf{0.986}	&0.974&	0.988&	0.875	&0.964	&0.856\\
&F1 score	&\textbf{0.958}	&0.836	&0.879&	0.544	&0.747&	0.428\\
\hline
\multirow{4}{*}{\makecell{New\\scenarios}}&MAE &\textbf{0.090}	&0.257&	0.299&	0.803&	0.943&	0.979 \\
&WMAPE	&\textbf{0.021}	&0.062&	0.071&	0.192&	0.226&	0.234\\
&Precision	&\textbf{1.000}	&0.879	&0.919	&0.824&	0.521&	0.867\\
&F1 score	&\textbf{0.863}	&0.827	&0.778	&0.503&	0.550&	0.382\\

\hline
\end{tabular}
}
\end{center}
\end{table}

\begin{figure*}[th]
\centering
    \subfloat[Cut in scenario]{\includegraphics[width = 0.98\textwidth]{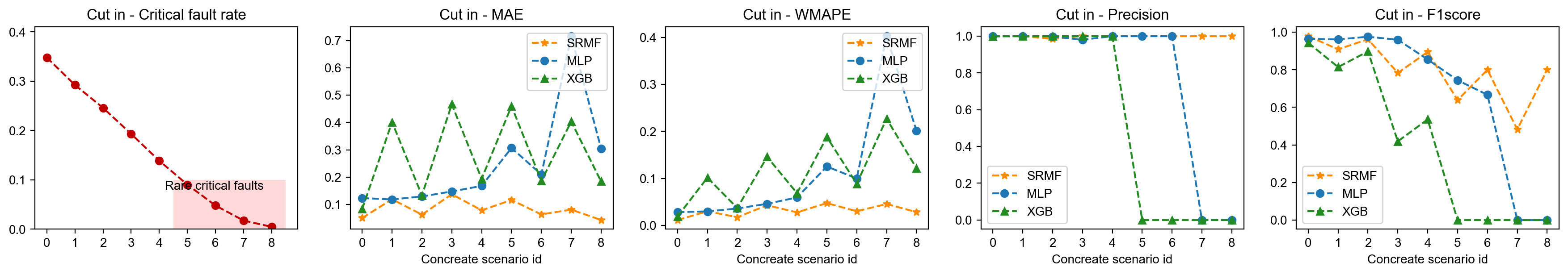}}
    \label{cut_in_metrics}
    \hfill
    \subfloat[Car following scenario]{\includegraphics[width = 0.98\textwidth]{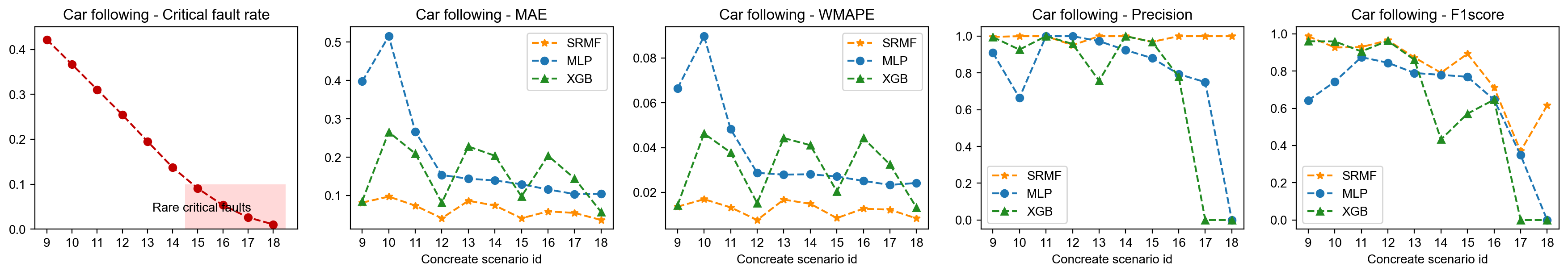}}
    \label{car_following_metrics}
  \caption{Comparison between \gls{strmf}, MLP and XGB in the cut in and car following scenario. In the first figure, the critical fault rate gradually decreases and the scenarios with rare critical faults are marked in red. The second to fifth figures are the MAE, WMAPE, Precision, F1 score of \gls{strmf}, MLP and XGB in different scenarios. The MAE, WMAPE of \gls{strmf} (colored orange) is smaller than MLP (colored blue) and XGB (colored green), and the Precision and F1 score of \gls{strmf} is higher than MLP and XGB overall.}
  \label{fig_metrics_xgb_and_strmf}
\end{figure*}

\gls{strmf} outperforms other surrogate models in all scenarios. It can be observed from Fig. \ref{simulation_and_prediction_results} and TABLE \ref{metrics_of_all_scenario}: 1) the prediction results of KNN, RBF and SVR are obviously different with the simulation results, with the largest prediction error. 2) \gls{strmf}, MLP and XGB can predict the feature of each scenarios basically, which is similar to simulation results in both the existing and new scenarios, the cut in and car following scenarios. 3) in the evaluation metrics of all scenarios, \gls{strmf} having the smallest MAE and WMAPE, the highest precision and F1 score. The precision can achieve $99.3\%$ and the F1 score is $91.1\%$.

In new scenarios, \gls{strmf} has the best generalization ability. To clearly compare the performance of these models in new scenarios, the prediction errors of existing and new scenarios are also shown in TABLE \ref{metrics_of_all_scenario}. The detailed metrics in each scenario of \gls{strmf}, MLP and XGB are displayed in Fig. \ref{fig_metrics_xgb_and_strmf}. It can be seen from Fig. \ref{fig_metrics_xgb_and_strmf} that \gls{strmf}, MLP and XGB perform better in existing scenarios than in new scenarios. However, MLP and XGB changes more dramatically and \gls{strmf} is more stable. The MAE of XGB in new scenarios increases by more than $103\%$ over the MAE in existing scenarios, which echo the shortcoming we conclude previously in the Introduction section that the prevailing methods overfit in the existing scenario but their generalizability is relatively low. Moreover, the precision and F1 score of \gls{strmf} can still achieve $100\%$ and $86.3\%$ in new scenarios.

\subsection{Effectiveness of \gls{strmf} in rare critical faults}
The critical fault rate is calculated by dividing the number of critical faults by the total number of fault types in this concrete scenario. As shown in Fig. \ref{fig_metrics_xgb_and_strmf}, the critical fault rate decreases as the initial distance increases. Here, we take $10\%$ as the threshold to define rare critical faults.

\begin{figure}[!thpb]
\centering
  \includegraphics[width=0.48\textwidth]{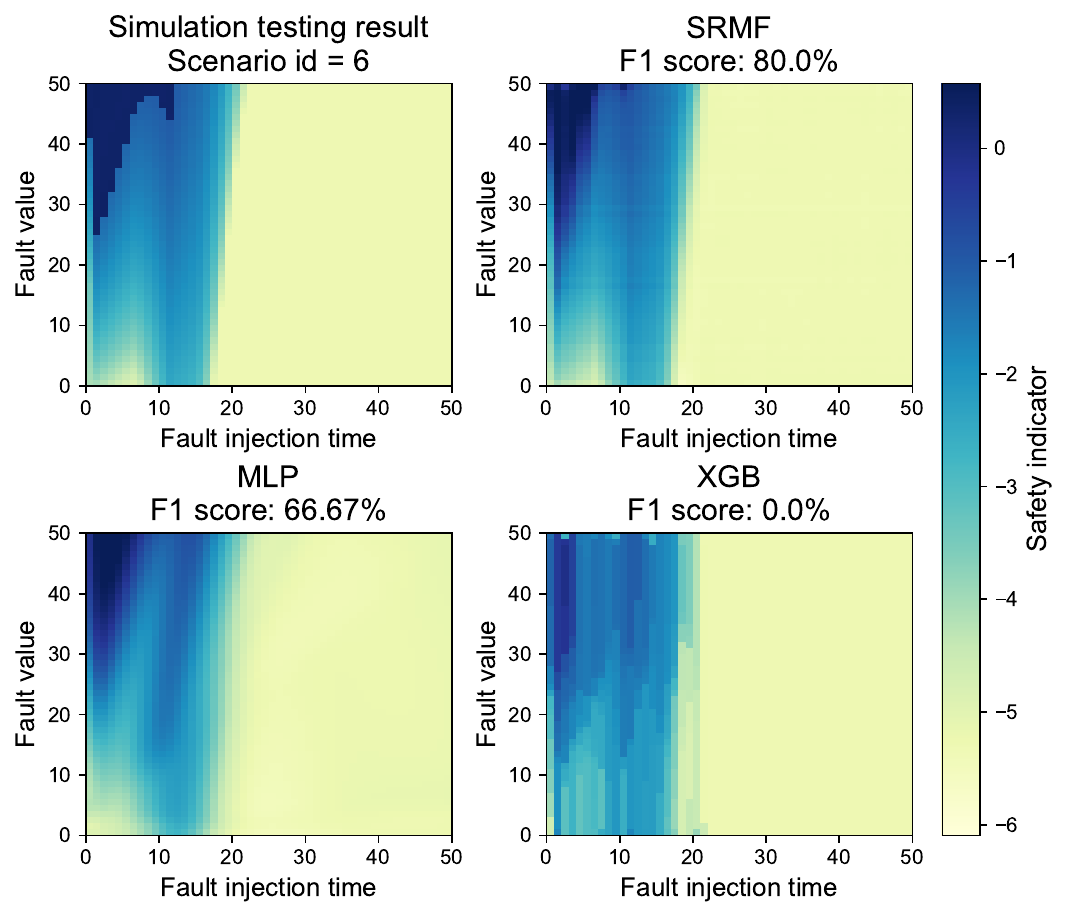}
  \caption{Simulation and prediction results of concrete scenario \# $6$.}
  \label{s_id_6}
\end{figure}

Taking scenario \# $6$ as an example, the simulation and prediction results are shown in Fig. \ref{s_id_6}. There are critical faults in the upper left (colored dark blue). \gls{strmf} also predicts these areas as critical faults with $80.0\%$ F1 score, while MLP with $66.67\%$ F1 score and XGB does not predict any critical faults.

TABLE \ref{rare_critical_faults} and Fig. \ref{fig_metrics_xgb_and_strmf} demonstrate the precision and F1 score of rare critical faults. In the cut in scenario, XGB fails to search for critical faults and the F1 score equals to $0$. MLP fails in scenario \# $7, 8$. However, \gls{strmf} still finds critical faults with $100\%$ precision and the F1 score can achieve $80.0\%$ in the two tested scenarios. In the car following scenario, XGB fails in scenario \# $17, 18$, but \gls{strmf} also succeeds in finding some rare critical faults.

\begin{table*}[htbp]
\caption{Performance in rare critical faults}
\vspace{-20pt}
\fontsize{7pt}{10pt}\selectfont
\label{rare_critical_faults}
\begin{center}
\begin{tabular}{cc|cccccccccccc}
\hline
\multirow{2}{*}{\makecell{Scenario\\id}} & \multirow{2}{*}{\makecell{Citical \\faults rate}} & \multicolumn{2}{c}{\gls{strmf}} & \multicolumn{2}{c}{MLP}& \multicolumn{2}{c}{XGB}& \multicolumn{2}{c}{KNN}& \multicolumn{2}{c}{RBF}& \multicolumn{2}{c}{SVR}\\
 & & Precison & F1 score & Precison & F1 score& Precison & F1 score& Precison & F1 score& Precison & F1 score& Precison & F1 score\\
\hline
5	&0.090&\textbf{1.000}&	0.639	&\textbf{1.000}&	\textbf{0.744}&	0.000	&0.000	&0.853	&0.505&	0.903&	0.706&	1.000&	0.101\\
6	&0.048&	\textbf{1.000}&	\textbf{0.800}&	\textbf{1.000}	&0.667&	0.000&	0.000	&0.775&	0.388&	0.984&	0.670	&1.000&	0.154\\
7	&0.018&\textbf{1.000}	&\textbf{0.483}&	0.000	&0.000&	0.000&	0.000&	0.577&	0.429	&0.889&	0.302&	0.667	&0.226\\
8&	0.005	&\textbf{1.000}	&\textbf{0.800}&	0.000	&0.000&	0.000&	0.000&	0.300&	0.375&	0.750	&0.375	&0.000	&0.000\\
\hline
15&	0.091&\textbf{0.969}	&\textbf{0.893}&	0.881	&0.769&	0.968	&0.571&	0.627	&0.535	&0.977&	0.546	&0.892	&0.617\\
16&	0.054&	\textbf{1.000}&	\textbf{0.712}&	0.793	&0.646&	0.779&	0.646&	0.449&	0.470	&0.892&	0.583	&0.609&	0.562\\
17&	0.026&	\textbf{1.000}	&\textbf{0.370}	&0.750	&0.349&	0.000&	0.000&	0.252	&0.338	&0.905	&0.704	&0.194	&0.253\\
18&	0.011&	\textbf{1.000}	&\textbf{0.615}	&0.000&	0.000	&0.000	&0.000	&0.103	&0.170	&0.714&	0.488	&0.013	&0.022\\
\hline
\end{tabular}
\end{center}
\end{table*}

\subsection{Parameter tuning and analysis}
\begin{figure}[!thpb]
\centering
  \includegraphics[width=0.48\textwidth]{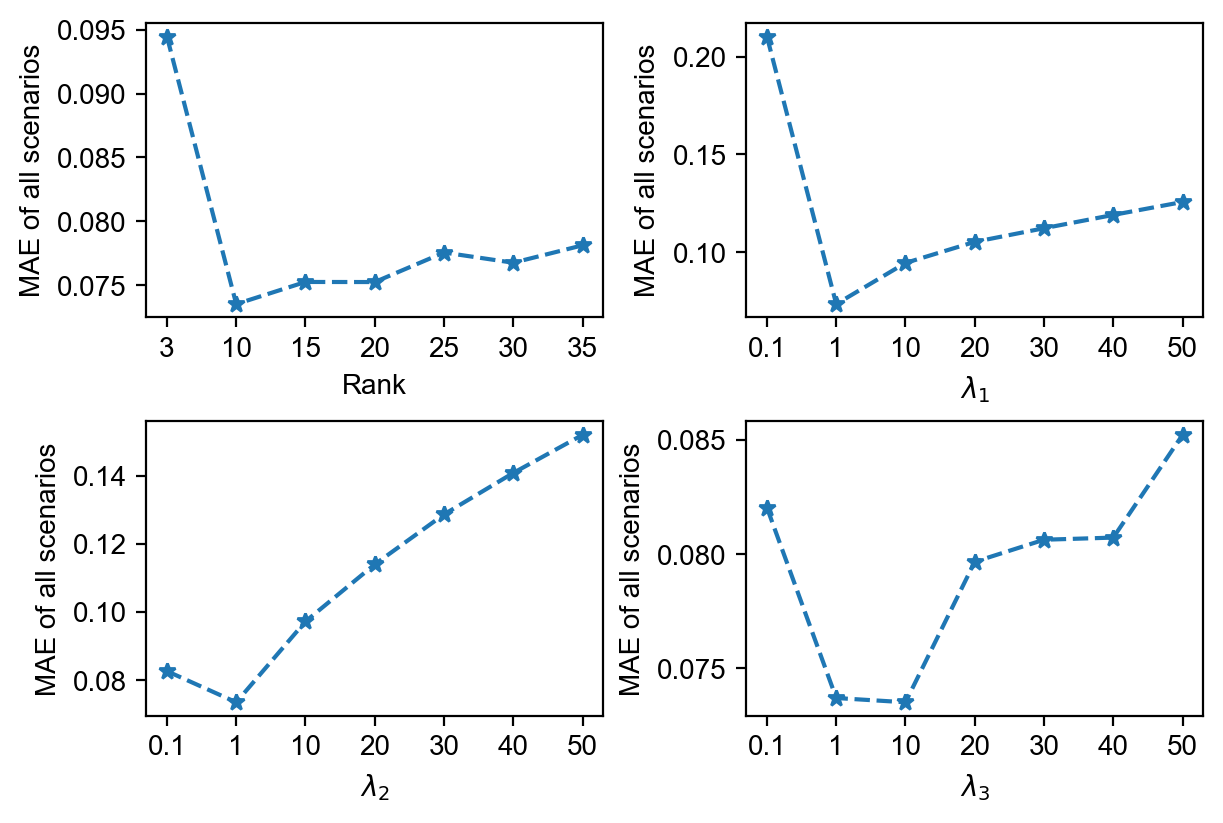}
  \caption{Model performance under different parameters: rank, $\lambda_1$, $\lambda_2$, $\lambda_3$.}
  \label{sensitivity}
\end{figure}

To analyze the influence of key hyper parameters on the performance of \gls{strmf}, we compare the model performance under different parameter settings, as shown in Fig. \ref{sensitivity}. The hyper parameters include the rank of the matrix, the coefficients of smoothness regularization $\lambda_1$, $\lambda_2$, $\lambda_3$. The MAE reaches its minimum when the rank is equal to 10 and tends to increase when the rank becomes greater than 10. For $\lambda$, \gls{strmf} has the smallest MAE when $\lambda_1$, $\lambda_2$ is set to 1 and when $\lambda_3$ equals 10. The MAE increases as the values of $\lambda_1$, $\lambda_2$, $\lambda_3$ decrease (equal to 0.1), demonstrating that smoothness regularization is useful for improving model performance.

\section{CONCLUSIONS}
Fault tolerance testing of HAVs is important and should be accelerated for the rapid development of HAVs due to the high-dimensional problem caused by numerous scenarios and faults. Surrogate models commonly employed to accelerate testing are prone to overfit in existing scenarios and might not detect rare critical faults. To overcome these problems, a new accelerated FI testing method based on \gls{strmf} is proposed. \gls{strmf} solves the high-dimensional problem by imposing the low-rank constraint and uses correlation between faults and scenarios to tackle the newly added scenario and the rare critical fault problem. 
It turns out that \gls{strmf} can perform well in new scenarios and is capable of identifying rare critical faults. The precision of \gls{strmf} is higher than 99\% and the acceleration rate can reach 1171 times.

There are also several directions to explore in future research. First, the details of \gls{strmf} can be improved to tackle larger dimensions with more types of scenarios and faults. Second, \gls{strmf}, which acts as a surrogate model, can be combined with other existing method in HAVs safety acceleration testing to find more rare critical faults.

\addtolength{\textheight}{12cm}   
\bibliographystyle{IEEEtran}
\bibliography{ref}

\begin{thebibliography}{10}
\providecommand{\url}[1]{#1}
\csname url@samestyle\endcsname
\providecommand{\newblock}{\relax}
\providecommand{\bibinfo}[2]{#2}
\providecommand{\BIBentrySTDinterwordspacing}{\spaceskip=0pt\relax}
\providecommand{\BIBentryALTinterwordstretchfactor}{4}
\providecommand{\BIBentryALTinterwordspacing}{\spaceskip=\fontdimen2\font plus
\BIBentryALTinterwordstretchfactor\fontdimen3\font minus \fontdimen4\font\relax}
\providecommand{\BIBforeignlanguage}[2]{{%
\expandafter\ifx\csname l@#1\endcsname\relax
\typeout{** WARNING: IEEEtran.bst: No hyphenation pattern has been}%
\typeout{** loaded for the language `#1'. Using the pattern for}%
\typeout{** the default language instead.}%
\else
\language=\csname l@#1\endcsname
\fi
#2}}
\providecommand{\BIBdecl}{\relax}
\BIBdecl

\bibitem{feng2023dense}
S.~Feng, H.~Sun, X.~Yan, H.~Zhu, Z.~Zou, S.~Shen, and H.~X. Liu, ``Dense reinforcement learning for safety validation of autonomous vehicles,'' \emph{Nature}, vol. 615, no. 7953, pp. 620--627, 2023.

\bibitem{hoque2022autonomous}
M.~A. Hoque and R.~Hasan, ``Autonomous driving security: A comprehensive threat model of attacks and mitigation strategies,'' in \emph{2022 IEEE 8th World Forum on Internet of Things (WF-IoT)}.\hskip 1em plus 0.5em minus 0.4em\relax IEEE, 2022, pp. 1--6.

\bibitem{jha2019kayotee}
S.~Jha, T.~Tsai, S.~Hari, M.~Sullivan, Z.~Kalbarczyk, S.~W. Keckler, and R.~K. Iyer, ``Kayotee: A fault injection-based system to assess the safety and reliability of autonomous vehicles to faults and errors,'' \emph{arXiv preprint arXiv:1907.01024}, 2019.

\bibitem{jha2020ml}
S.~Jha, S.~Cui, S.~Banerjee, J.~Cyriac, T.~Tsai, Z.~Kalbarczyk, and R.~K. Iyer, ``Ml-driven malware that targets av safety,'' in \emph{2020 50th annual IEEE/IFIP international conference on dependable systems and networks (DSN)}.\hskip 1em plus 0.5em minus 0.4em\relax IEEE, 2020, pp. 113--124.

\bibitem{ma_verification_2022}
Y.~Ma, C.~Sun, J.~Chen, D.~Cao, and L.~Xiong, ``Verification and {Validation} {Methods} for {Decision}-{Making} and {Planning} of {Automated} {Vehicles}: {A} {Review},'' \emph{IEEE Transactions on Intelligent Vehicles}, vol.~7, no.~3, pp. 480--498, Sep. 2022.

\bibitem{Sun2022scenario-based}
J.~Sun, H.~Zhang, H.~Zhou, R.~Yu, and Y.~Tian, ``Scenario-based test automation for highly automated vehicles: A review and paving the way for systematic safety assurance,'' \emph{IEEE Transactions on Intelligent Transportation Systems}, vol.~23, no.~9, pp. 14\,088--14\,103, 2022.

\bibitem{chen_adaptive_2023}
Q.~Chen, H.~Zhang, H.~Zhou, J.~Sun, and Y.~Tian, ``Adaptive {Design} of {Experiments} for {Fault} {Injection} {Testing} of {Highly} {Automated} {Vehicles},'' \emph{IEEE Intelligent Transportation Systems Magazine}, pp. 2--19, 2023.

\bibitem{sun2021adaptive}
J.~Sun, H.~Zhou, H.~Xi, H.~Zhang, and Y.~Tian, ``Adaptive design of experiments for safety evaluation of automated vehicles,'' \emph{IEEE Transactions on Intelligent Transportation Systems}, vol.~23, no.~9, pp. 14\,497--14\,508, 2021.

\bibitem{liu2012tensor}
J.~Liu, P.~Musialski, P.~Wonka, and J.~Ye, ``Tensor completion for estimating missing values in visual data,'' \emph{IEEE Transactions on Pattern Analysis and Machine Intelligence}, vol.~35, no.~1, pp. 208--220, 2012.

\bibitem{deng2022new}
L.~Deng and M.~Xiao, ``A new automatic hyperparameter recommendation approach under low-rank tensor completion e framework,'' \emph{IEEE Transactions on Pattern Analysis and Machine Intelligence}, vol.~45, no.~4, pp. 4038--4050, 2022.

\bibitem{nie2023correlating}
T.~Nie, G.~Qin, Y.~Wang, and J.~Sun, ``Correlating sparse sensing for large-scale traffic speed estimation: A laplacian-enhanced low-rank tensor kriging approach,'' \emph{Transportation Research Part C: Emerging Technologies}, vol. 152, p. 104190, 2023.

\bibitem{Moradi2020}
M.~Moradi, B.~J. Oakes, M.~Saraoglu, A.~Morozov, K.~Janschek, and J.~Denil, ``Exploring fault parameter space using reinforcement learning-based fault injection,'' in \emph{2020 50th Annual IEEE/IFIP International Conference on Dependable Systems and Networks Workshops (DSN-W)}, 2020, pp. 102--109.

\bibitem{mei2024}
Y.~Mei, T.~Nie, J.~Sun, and Y.~Tian, ``Bayesian fault injection safety testing for highly automated vehicles with uncertainty,'' \emph{IEEE Transactions on Intelligent Vehicles}, pp. 1--15, 2024.

\bibitem{Beglerovic2017}
H.~Beglerovic, M.~Stolz, and M.~Horn, ``Testing of autonomous vehicles using surrogate models and stochastic optimization,'' in \emph{2017 IEEE 20th International Conference on Intelligent Transportation Systems (ITSC)}, 2017, pp. 1--6.

\bibitem{MULLINS2018197}
\BIBentryALTinterwordspacing
G.~E. Mullins, P.~G. Stankiewicz, R.~C. Hawthorne, and S.~K. Gupta, ``Adaptive generation of challenging scenarios for testing and evaluation of autonomous vehicles,'' \emph{Journal of Systems and Software}, vol. 137, pp. 197--215, 2018. [Online]. Available: \url{https://www.sciencedirect.com/science/article/pii/S0164121217302546}
\BIBentrySTDinterwordspacing

\bibitem{wang2022safety}
Y.~Wang, R.~Yu, S.~Qiu, J.~Sun, and H.~Farah, ``Safety performance boundary identification of highly automated vehicles: A surrogate model-based gradient descent searching approach,'' \emph{IEEE Transactions on Intelligent Transportation Systems}, vol.~23, no.~12, pp. 23\,809--23\,820, 2022.

\bibitem{angione2022using}
C.~Angione, E.~Silverman, and E.~Yaneske, ``Using machine learning as a surrogate model for agent-based simulations,'' \emph{Plos one}, vol.~17, no.~2, p. e0263150, 2022.

\bibitem{tan2013tensor}
H.~Tan, G.~Feng, J.~Feng, W.~Wang, Y.-J. Zhang, and F.~Li, ``A tensor-based method for missing traffic data completion,'' \emph{Transportation Research Part C: Emerging Technologies}, vol.~28, pp. 15--27, 2013.

\bibitem{nie2022truncated}
T.~Nie, G.~Qin, and J.~Sun, ``Truncated tensor schatten p-norm based approach for spatiotemporal traffic data imputation with complicated missing patterns,'' \emph{Transportation Research Part C: Emerging Technologies}, vol. 141, p. 103737, 2022.

\bibitem{asif2016matrix}
M.~T. Asif, N.~Mitrovic, J.~Dauwels, and P.~Jaillet, ``Matrix and tensor based methods for missing data estimation in large traffic networks,'' \emph{IEEE Transactions on Intelligent Transportation Systems}, vol.~17, no.~7, pp. 1816--1825, 2016.

\bibitem{yang2021real}
J.-M. Yang, Z.-R. Peng, and L.~Lin, ``Real-time spatiotemporal prediction and imputation of traffic status based on lstm and graph laplacian regularized matrix factorization,'' \emph{Transportation Research Part C: Emerging Technologies}, vol. 129, p. 103228, 2021.

\bibitem{nie2023imputeformer}
T.~Nie, G.~Qin, W.~Ma, Y.~Mei, and J.~Sun, ``Imputeformer: Low rankness-induced transformers for generalizable spatiotemporal imputation,'' \emph{arXiv: 2312.01728}, 2023.

\bibitem{nie2024spatiotemporal}
T.~Nie, G.~Qin, W.~Ma, and J.~Sun, ``Spatiotemporal implicit neural representation as a generalized traffic data learner,'' \emph{arXiv preprint arXiv:2405.03185}, 2024.

\bibitem{luise2019leveraging}
G.~Luise, D.~Stamos, M.~Pontil, and C.~Ciliberto, ``Leveraging low-rank relations between surrogate tasks in structured prediction,'' in \emph{International Conference on Machine Learning}.\hskip 1em plus 0.5em minus 0.4em\relax PMLR, 2019, pp. 4193--4202.

\bibitem{PEGASUS2017Scenario}
\BIBentryALTinterwordspacing
PEGASUS. Scenario description. [Online]. Available: \url{https://www.pegasusprojekt.de/files/tmpl/PDFSymposium/04_Scenario-Description.pdf}
\BIBentrySTDinterwordspacing

\bibitem{xing_adaptive_2023}
X.~Xing, L.~Liu, J.~Chen, L.~Xiong, Y.~Huang, and Z.~Yu, ``Adaptive error injection for robustness verification of decision-making systems for autonomous vehicles,'' \emph{Proceedings of the Institution of Mechanical Engineers, Part D: Journal of Automobile Engineering}, 2023.

\bibitem{Jha2019ML-Based}
S.~Jha, S.~Banerjee, T.~Tsai, S.~K. Hari, M.~B. Sullivan, Z.~T. Kalbarczyk, S.~W. Keckler, and R.~K. Iyer, ``Ml-based fault injection for autonomous vehicles: A case for bayesian fault injection,'' in \emph{2019 49th annual IEEE/IFIP international conference on dependable systems and networks (DSN)}.\hskip 1em plus 0.5em minus 0.4em\relax IEEE, 2019, pp. 112--124.

\bibitem{yu2016temporal}
H.-F. Yu, N.~Rao, and I.~S. Dhillon, ``Temporal regularized matrix factorization for high-dimensional time series prediction,'' \emph{Advances in neural information processing systems}, vol.~29, 2016.

\bibitem{chen2022nonstationary}
X.~Chen, C.~Zhang, X.-L. Zhao, N.~Saunier, and L.~Sun, ``Nonstationary temporal matrix factorization for multivariate time series forecasting,'' \emph{arXiv preprint arXiv:2203.10651}, 2022.

\end{thebibliography}

\end{document}